\pdfoutput=1
\documentclass[manuscript, nonacm]{acmart}

\renewcommand\footnotetextcopyrightpermission[1]{}
\begin{document}

\title{Generative UI as an Accessibility Bridge: Lessons from C2C E-Commerce}

\author{Bektur Ryskeldiev}
\affiliation{%
  \institution{Mercari R4D}
  \city{Tokyo}
  \country{Japan}}
\affiliation{%
  \institution{University of Tsukuba}
  \city{Tsukuba}
  \country{Japan}}
\email{bektour@mercari.com}

\begin{abstract}
Web accessibility rests on static standards and developer compliance. That model frays in platforms where content is user-generated: photos arrive blurry or off-frame, descriptions skip size and condition, and page structure shifts from listing to listing. Drawing on six studies conducted between 2022 and 2025 with blind, low-vision, and older adult users of customer-to-customer (C2C) marketplaces, I argue that generative UI can produce adapted interfaces at the point of use, addressing barriers that static design cannot anticipate. Three interventions from this program---HTML regeneration for screen readers, conversational guidance for older sellers, and audio-guided photo framing for blind sellers---demonstrate how runtime generation can bridge gaps that standards leave open. I outline what these findings imply for HCI practice: generative UI extends beyond the screen, complements rather than replaces ability-based design, and shifts the designer's role from specifying layouts to specifying policies. This is an expanded arXiv version of a position paper accepted at the CHI 2026 workshop \emph{What does Generative UI mean for HCI Practice?}
\end{abstract}


\begin{teaserfigure}
  \includegraphics[width=\textwidth]{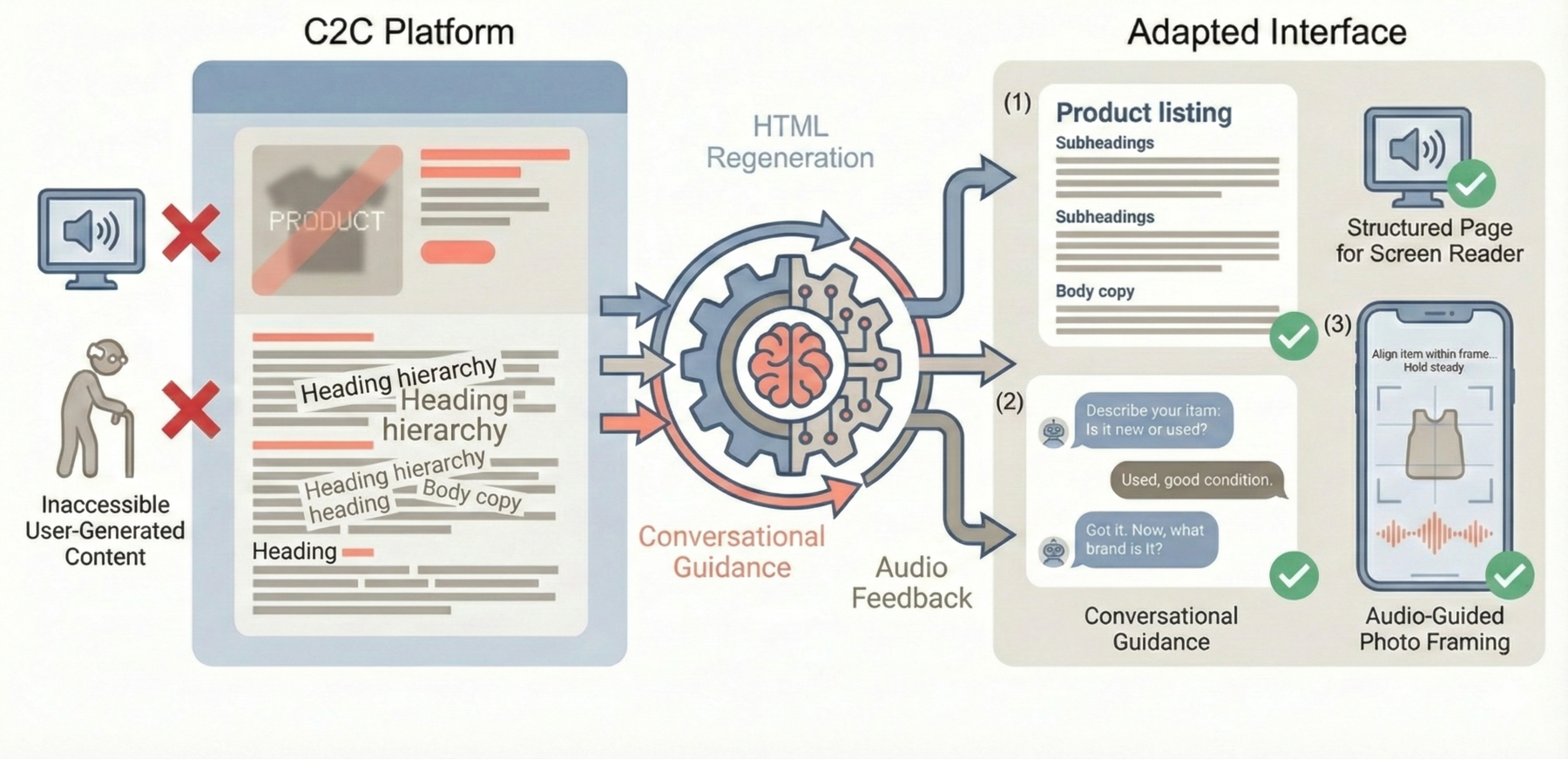}
  \caption{Generative UI as an accessibility bridge in C2C e-commerce. Inaccessible user-generated content (left) is transformed through three GenUI interventions: (1) HTML regeneration for screen readers, (2) conversational guidance for step-by-step listing, and (3) audio-guided photo framing for blind sellers.}
  \Description{A diagram showing how a C2C platform with cluttered, inaccessible content is transformed through an AI layer into three adapted interfaces: a structured page for screen readers, a conversational chatbot for listing guidance, and a mobile camera with audio feedback for product photography.}
\end{teaserfigure}

\maketitle

\section{Introduction}

Accessibility standards such as WCAG~\cite{w3c2018wcag} assume a clean division of labor: developers build compliant interfaces, users consume them. That division holds up on platforms where content is authored centrally. It breaks down in customer-to-customer (C2C) marketplaces, where individuals list to each other. Product photos are taken by amateurs, item descriptions skip color and condition, and page structure drifts from listing to listing. No alt-text policy can guarantee what a seller uploaded.

Between 2022 and 2025, my collaborators and I ran six studies with blind, low-vision, and older adult users of C2C platforms~\cite{ryskeldiev2022investigating, ryskeldiev2023immersive, park2024conversational, yu2025cluttered, gupta2025accessibleship, tsutsui2025blind}. A consistent pattern emerged across them: accessibility barriers persisted despite platform-level WCAG compliance, because the problems originated in content no designer could fully control. Generative UI---interfaces produced by AI models at runtime~\cite{lindley2026genui}---offers a way to close these gaps by reshaping the interface at the moment of use, adapting to both content and user.

This paper synthesizes that body of work into three claims. First, static accessibility standards cannot reach user-generated content; a second mechanism is needed at render time. Second, three recent prototypes from our program---an HTML-regenerating browser extension, a chatbot listing assistant, and an audio-guided photography tool---demonstrate that runtime GenUI can close specific C2C barriers, with measurable improvements over existing assistive tools. Third, these results imply a shift for HCI practice: from specifying layouts to specifying the policies and constraints under which interfaces are generated.

The original two-page position paper, accepted at the CHI 2026 workshop on Generative UI, presented these claims in compressed form. This arXiv version adds a related work section, expanded quantitative detail on each intervention, and a discussion of limitations and open questions.

\section{Related Work}
\label{sec:related}

\subsection{Accessibility standards and their reach}

WCAG~\cite{w3c2018wcag} has been the dominant accessibility reference since 2008, codifying requirements for perceivability, operability, understandability, and robustness. These guidelines target what developers ship. They are silent on what sellers upload. Prior e-commerce accessibility work~\cite{liu2019ordinary, stangl2018browsewithme, wang2021revamp} documents repeated failures on product pages even when sites pass automated audits: missing alt text, inconsistent headings, controls labeled only as ``button.'' WCAG pass rates correlate poorly with what a screen reader user actually hears.

\subsection{Adaptive and ability-based interfaces}

A separate line of work---originating in SUPPLE~\cite{gajos2004supple, gajos2007supple++} and formalized as \emph{ability-based design} by Wobbrock et al.~\cite{wobbrock2011ability}---treats the interface itself as adaptable. SUPPLE used constraint optimization to render forms suited to a user's motor and vision capabilities. Ability-based design generalized this to a design stance: start from what the user can do, and let systems and interfaces adapt. These systems assume structured, known inputs: form fields, menu items, discrete widgets.

Generative UI extends this lineage in one direction and weakens an assumption in another. It extends ability-based design by adapting to arbitrary content---a seller's photo, a description with missing attributes---not only to the user's capabilities. It weakens the assumption that the adaptation target is structured: a large language model can restructure a page's HTML, narrate a scene, or rewrite a description on demand.

\subsection{Generative UI as an emerging paradigm}

The framing of generative UI as an HCI problem is recent~\cite{lindley2026genui}. Where classical UI generation targeted layout under constraints, contemporary GenUI additionally targets content-aware transformation: the interface is generated from the underlying content, the user's context, and a task, rather than specified in advance. This shifts what designers control---from pixels to prompts---and raises questions this paper tries to sharpen: where does GenUI actually help, and where does it fail, for users whom static accessibility has underserved?

\subsection{Accessibility of online shopping}

C2C accessibility has been studied less than B2C accessibility, but the difference matters. Nagatani et al.~\cite{nagatani2022mediause} report that 93\% of blind and 96\% of low-vision respondents in Japan shop online at least occasionally, yet far fewer sell online. Prior systems have improved the buying side: BrowseWithMe~\cite{stangl2018browsewithme} conversationalizes clothing catalogs; Revamp~\cite{wang2021revamp} extracts and reorganizes product pages. Image description preferences have been studied in depth~\cite{stangl2020persontree}. The selling side---listing, photographing, packaging, shipping---has received far less attention, and is where C2C accessibility barriers concentrate.

\section{Where Static Accessibility Falls Short}

Our foundational ASSETS~'22 study~\cite{ryskeldiev2022investigating} interviewed twelve low-vision participants in Japan using a working C2C platform. Three patterns emerged that WCAG compliance cannot address.

\paragraph{User-generated content carries user-generated defects.} Eight of the twelve participants reported difficulty with seller-uploaded product photos---low contrast, motion blur, item colors matching the background---and with descriptions missing color, condition, or size information. No platform-level alt-text policy guarantees the information a casual seller chose not to write down.

\paragraph{Accessibility barriers extend beyond the screen.} Six participants who also sold items struggled with product photography, and eight reported difficulty with packaging---choosing an appropriately sized box, confirming an item's condition, filling in a handwritten shipping slip. These tasks sit outside the software interface entirely, but determine whether someone can participate as a seller at all. A later CHI~EA~'25 study~\cite{gupta2025accessibleship} made this explicit: up to 70\% of people with visual impairments refrain from selling on C2C platforms, a figure originally reported by Nagatani et al.~\cite{nagatani2022mediause}.

\paragraph{Structural inaccessibility compounds as pages scale.} In a later ASSETS~'25 study~\cite{yu2025cluttered}, fifteen screen reader users spent significant time parsing cluttered HTML on Amazon, Nordstrom, and Mercari---heading hierarchies misused for layout, category menus presented as ten-level nested headings, reviews placed before product titles in reading order. Participant P3 summarized the state succinctly: \emph{``Amazon is an insanely cluttered website. There are so many headings. The only way to navigate Amazon is to just arrow down each item instead of using headings.''} Static audits catch few of these problems, because the HTML technically validates.

Across these studies, the same shape of problem recurs: the \emph{rendered experience} diverges from what the \emph{specification} would predict, because both content and structure are partly outside the platform's control.

\section{Three GenUI Interventions}

Three prototypes from our program illustrate how runtime generation can close the gaps left by static standards. Each targets a different point in the C2C journey: reading for buyers, writing for older sellers, photographing for blind sellers.

\subsection{HTML regeneration for screen readers}

Our ASSETS~'25 browser extension~\cite{yu2025cluttered} uses GPT-4o~\cite{openai2024gpt4o} to restructure e-commerce HTML for screen reader users. The system operates in two modes. \emph{Option 1} produces a fully regenerated, text-only HTML document optimized for screen reader navigation: headings re-ordered into a linear reading sequence, summary headings inserted, single-category lists stripped. \emph{Option 2} performs tag-level reorganization without altering visual layout: only the tag structure, ARIA attributes, and labels change.

Fifteen screen reader users evaluated both options against the original Mercari site. The regenerated version scored a mean of 5.0 on a 5-point Likert for overall browsing experience, compared with 4.57 for the tag-reorganized version and 3.14 for the original. Task completion time dropped correspondingly---from a median near 130 seconds on the original to near 25 seconds on Option 1. Automated audits (Lighthouse, SortSite, AChecker) showed reductions in WCAG Level A violations across all three tested sites, and aggregated semantic similarity between original and regenerated pages averaged 96.3\% (range 91.60\%--99.36\%), indicating that content meaning was largely preserved. The LLM produced a new interface from existing content, tailored to how screen reader users actually navigate.

\subsection{Conversational listing for older adults}

An ASSETS~'24 prototype~\cite{park2024conversational} guided ten older adult participants (aged 65--76) through listing items for sale via a rule-based conversational chatbot. The chatbot prompted users step by step: item name, category, description, price, photo, confirmation. All ten participants completed a listing.

What the study revealed was narrower than usability. Participants who struggled with understanding how separate listing tasks related to each other found the step-by-step conversational format easier to follow, because it leveraged a modality already familiar from daily messaging apps. Five participants reported that taking photos felt effortless because ``we usually send images through KakaoTalk'' (P4). The intervention did not solve the literacy barrier by adding features; it solved it by re-rendering the same task in a familiar conversational shape. Older adults did not need a different listing flow. They needed the same flow delivered by a different interface.

\subsection{AI-assisted product photography}

Our CHI~EA~'25 prototype~\cite{tsutsui2025blind} combines real-time object detection (COCO-SSD) with GPT-4o mini to help blind sellers take product photos. The capture screen runs object detection every 100 milliseconds, announces the item's position relative to a 3$\times$3 frame grid via audio feedback, and triggers capture when the user taps once the item is centered. After capture, the user can query the photo in natural language---\emph{``what color is it?''}, \emph{``is there a label visible?''}---and receive a voice response.

Four blind or low-vision participants compared the prototype with the iPad's default camera (with VoiceOver) and with Seeing~AI. Centering accuracy, measured as pixel offset from image center, was 46.49 pixels for the prototype, 122.99 for Seeing~AI, and 127.49 for the iPad camera. System Usability Scale scores were 73.12 (prototype), 70.0 (Seeing~AI), and 56.25 (iPad camera)---adjectivally ``Good,'' ``Good,'' and ``Poor.'' One participant framed the qualitative difference: \emph{``I used to have my parents check the photos I took, but now I feel like I can do it by myself.''} The generated interface was not visual. It was a stream of positional audio followed by a conversational image analysis.

\section{Implications for GenUI and HCI Practice}

\subsection{GenUI fills gaps standards cannot}

WCAG targets developer behavior. GenUI targets the rendered experience and adapts at the point of consumption. The distinction matters most where content quality is outside platform control---user-generated listings, amateur photos, seller-authored descriptions. In the HTML regeneration study, every test page passed some automated audits while still failing real users. The gap is not between compliant and non-compliant sites. It is between a site's code and the experience a screen reader delivers. GenUI operates in that gap.

\subsection{GenUI extends beyond the screen}

Our work on packaging~\cite{gupta2025accessibleship} and photography~\cite{tsutsui2025blind} shows that C2C accessibility barriers span digital and physical tasks. A seller cannot list without photographing, cannot ship without packaging, cannot respond to a buyer without tracking a delivery. The ``interface'' that GenUI generates need not be visual or even on-screen. In the photography prototype, the generated interface was a real-time audio stream describing where the object sat relative to a frame. Designing GenUI for accessibility means designing across modalities---conversational, auditory, multimodal---not just restructuring DOM trees.

\subsection{Ability-based design meets generative systems}

SUPPLE~\cite{gajos2004supple, gajos2007supple++} and ability-based design~\cite{wobbrock2011ability} adapted interfaces to the user's motor and vision capabilities, given structured inputs. GenUI extends this lineage by adding the ability to restructure arbitrary, loosely-structured content---HTML written by a third party, a seller's photo, a free-text description---based on user needs. The two approaches are complementary, not competing. Ability-based design supplies the principled grounding (know the user's capabilities, adapt to them) that GenUI on its own lacks. GenUI supplies the flexibility that ability-based design, built for structured inputs, could not reach into user-generated content.

\subsection{Evaluation needs both content fidelity and adaptation quality}

The HTML regeneration study balanced three evaluation lenses: automated accessibility audits (Lighthouse, SortSite, AChecker), aggregated semantic similarity between original and generated pages, and qualitative user evaluation. No one metric captures the goal. A regenerated page that scores 100 on Lighthouse but drops 30\% of the original product attributes is worse than the page it replaced. A regenerated page with perfect semantic similarity but unchanged heading hierarchy solves nothing. GenUI evaluation needs to measure \emph{fidelity to the source} and \emph{quality of the adaptation} simultaneously, and accept that these can trade off. In practice we set a semantic similarity threshold of 90\% as a floor before deploying any generated output.

\subsection{Design practice shifts from layouts to policies}

When GenUI adapts content at runtime, designers no longer control the final interface directly. What they control instead is the prompt, the constraints, the acceptance thresholds, and the fallbacks. The task becomes specifying \emph{generation policies}: which elements must be preserved (product price, shipping option), which can be restructured (heading hierarchy, section order), which require a user confirmation before replacing, and what the system should do when generation fails. This is closer to designing an API than designing a screen. It also raises questions about accountability---who is responsible when a generated interface misrepresents a seller's item---that static design largely side-stepped.

\section{Limitations and Open Questions}

\subsection{Content integrity and hallucination}

Large language models fabricate. In the HTML regeneration system, we observed the model occasionally condensing fragmented promotional elements into coherent sections---helpful in most cases, but once it dropped a link to an extended-warranty option on an Amazon product page. We addressed this with a semantic similarity threshold and targeted re-insertion of product links, but the underlying risk remains: GenUI for accessibility is GenUI mediating between a seller and a buyer who may never meet, and the cost of a subtly wrong description falls on both.

\subsection{Latency and cost}

Full-page HTML regeneration took one to five minutes per page in the ASSETS~'25 study and consumed roughly 220{,}000 tokens per session. For a screen reader user browsing quickly, minute-scale latency is a regression, not an improvement. Practical deployment requires caching, smaller specialized models, or regeneration only on first visit. The photography tool had the opposite problem: object detection ran every 100~ms to provide responsive audio framing, at the cost of battery life and thermal headroom on mobile devices. Neither constraint is fundamental, but both rule out naive deployments.

\subsection{Evaluation beyond convenience samples}

The studies surveyed here have sample sizes of 8 to 15. That is adequate for qualitative work and for detecting large effects, but not for the longitudinal question we have not yet answered: does a GenUI-adapted interface change whether someone sells at all? The 70\% refrain-from-selling figure~\cite{nagatani2022mediause, gupta2025accessibleship} is a population-level number that our interventions have not yet been evaluated against. Longer-term deployment studies are needed.

\subsection{Cultural and linguistic scope}

All our studies were conducted in Japan or the United States, and all with English or Japanese content. Generative UI depends on the underlying model's performance, and that performance varies by language and domain. What holds for Mercari may or may not hold for other C2C platforms, and what holds for GPT-4o may not hold for smaller or multilingual models.

\subsection{Trust, transparency, and user control}

Participants in the HTML regeneration study uniformly preferred the regenerated version, but only after they were told it was generated. Some asked about how to revert, how to know what changed, and whether their choices were tracked. GenUI that restructures a page on the user's behalf inherits the trust problems of content moderation: invisible changes are harder to consent to than visible ones. Design patterns for transparency---change summaries, toggle to original, provenance indicators---are an open area.

\section{Conclusion}

Accessibility standards describe what a compliant developer should build. They do not describe what a user with a screen reader actually hears, or what a blind seller holds in their hands before the shutter clicks. Across six studies in C2C marketplaces, the gap between the two has been the source of most real barriers. Generative UI, used carefully, can close that gap---not by replacing standards, but by operating at runtime where standards cannot reach. The three interventions presented here are early. They suggest that the more interesting question for HCI practice is no longer how to make a page compliant, but how to specify the policies under which a generated page remains both faithful and accessible.

\begin{acks}
This work builds on research conducted at Mercari R4D with collaborators at Mercari Inc., the University of Tsukuba (Digital Nature Group), the University of Illinois Urbana-Champaign, Singapore Management University, the University of Auckland, the University of Tokyo, Kyung Hee University, Seoul National University, and the University of Seoul. I thank the participants of the six studies discussed here, and the organizers and attendees of the CHI 2026 workshop on Generative UI for the conversations that shaped this paper.
\end{acks}

%%
%% Bibliography
%%
\bibliographystyle{ACM-Reference-Format}
\bibliography{genui-position-paper}

\end{document}